\documentclass[10pt]{article}

\usepackage[utf8]{inputenc}

\pdfoutput=1

\usepackage[oldcommands,boxed]{algorithm2e}

\usepackage[protrusion=true,expansion]{microtype} 
\usepackage[usenames,dvipsnames]{color}
\usepackage{amsmath, graphicx, latexsym, amssymb, amsthm, amsfonts}
\usepackage[mathscr]{eucal}
\usepackage{arydshln}
\usepackage{color}
\usepackage{epsfig}
\usepackage{url}
\usepackage{hyperref}
\usepackage{psfrag}
\usepackage{empheq}

\usepackage{pdflscape}

\usepackage{wrapfig}

\usepackage[update,prepend]{epstopdf}
\usepackage{wasysym}

\usepackage[subnum]{cases}

\usepackage[usenames,dvipsnames]{color}
\usepackage{amsmath, graphicx, latexsym, amssymb, amsthm, amsfonts}
\usepackage[mathscr]{eucal}
\usepackage{graphics}\usepackage{graphicx}
\usepackage{color}
\usepackage{epsfig}
\usepackage{hyperref}
\newcommand{\TT}{{\mathbb T}}
\newcommand{\T}{{\mathcal T}}

\newcommand{\MM}{{\mathbb M}}

\newcommand{\SSS}{{\mathbb S}}

\newcommand{\AAA}{{\mathbb A}}
\newcommand{\GG}{{\mathbb G}}
\newcommand{\CC}{{\mathbb C}}
\newcommand{\LL}{{\mathbb L}}
\newcommand{\HH}{{\mathbb H}}

\newcommand{\bI}{ {\bf I} }

\begin{document}

\newtheorem{lemma}{Lemma}
\newtheorem{theorem}{Theorem}
\newtheorem{corollary}{Corollary}
\newtheorem{definition}{Definition}
\newtheorem{example}{Example}
\newtheorem{proposition}{Proposition}
\newtheorem{condition}{Condition}
\newtheorem{assumption}{Assumption}
\newtheorem{conjecture}{Conjecture}
\newtheorem{problem}{Problem}
\newtheorem{remark}{Remark}

\def\thelemma{\arabic{section}.\arabic{lemma}}
\def\thetheorem{\arabic{section}.\arabic{theorem}}
\def\thecorollary{\arabic{section}.\arabic{corollary}}
\def\thedefinition{\arabic{section}.\arabic{definition}}
\def\theexample{\arabic{section}.\arabic{example}}
\def\theproposition{\arabic{section}.\arabic{proposition}}
\def\thecondition{\arabic{section}.\arabic{condition}}
\def\theassumption{\arabic{section}.\arabic{assumption}}
\def\theconjecture{\arabic{section}.\arabic{conjecture}}
\def\theproblem{\arabic{section}.\arabic{problem}}
\def\theremark{\arabic{section}.\arabic{remark}}

\newcommand{\manualnames}[1]{
\def\thelemma{#1.\arabic{lemma}}
\def\thetheorem{#1.\arabic{theorem}}
\def\thecorollary{#1.\arabic{corollary}}
\def\thedefinition{#1.\arabic{definition}}
\def\theexample{#1.\arabic{example}}
\def\theproposition{#1.\arabic{proposition}}
\def\theassumption{#1.\arabic{assumption}}
\def\theremark{#1.\arabic{remark}}
}

\newcommand{\beginsec}{
\setcounter{lemma}{0}
\setcounter{theorem}{0}
\setcounter{corollary}{0}
\setcounter{definition}{0}
\setcounter{example}{0}
\setcounter{proposition}{0}
\setcounter{condition}{0}
\setcounter{assumption}{0}
\setcounter{conjecture}{0}
\setcounter{problem}{0}
\setcounter{remark}{0}
}
\newcommand{\la}{\lambda}
\newcommand{\eps}{\varepsilon}
\newcommand{\ph}{\varphi}
\newcommand{\vr}{\varrho}
\newcommand{\al}{\alpha}
\newcommand{\bet}{\beta}
\newcommand{\gam}{\gamma}
\newcommand{\kap}{\kappa}
\newcommand{\s}{\sigma}
\newcommand{\sig}{\sigma}
\newcommand{\om}{\omega}
\newcommand{\Gam}{\mathnormal{\Gamma}}
\newcommand{\Del}{\mathnormal{\Delta}}
\newcommand{\Th}{\mathnormal{\Theta}}
\newcommand{\La}{\mathnormal{\Lambda}}
\newcommand{\X}{\mathnormal{\Xi}}
\newcommand{\PI}{\mathnormal{\Pi}}
\newcommand{\Sig}{\mathnormal{\Sigma}}
\newcommand{\Ups}{\mathnormal{\Upsilon}}
\newcommand{\Ph}{\mathnormal{\Phi}}
\newcommand{\Ps}{\mathnormal{\Psi}}
\newcommand{\Om}{\mathnormal{\Omega}}

\newcommand{\EE}{{\mathbb E}}
\newcommand{\FF}{{\mathbb F}}
\newcommand{\PP}{{\mathbb P}}
\newcommand{\ONE}{\boldsymbol{1}}

\newcommand{\calA}{{\cal A}}
\newcommand{\calB}{{\cal B}}
\newcommand{\calC}{{\cal C}}
\newcommand{\calD}{{\cal D}}
\newcommand{\calE}{{\cal E}}
\newcommand{\calF}{{\cal F}}
\newcommand{\calG}{{\cal G}}
\newcommand{\calH}{{\cal H}}
\newcommand{\calI}{{\cal I}}
\newcommand{\calJ}{{\cal J}}
\newcommand{\calL}{{\cal L}}
\newcommand{\calM}{{\cal M}}
\newcommand{\calN}{{\cal N}}
\newcommand{\calP}{{\cal P}}
\newcommand{\calR}{{\cal R}}
\newcommand{\calS}{{\cal S}}
\newcommand{\calT}{{\cal T}}
\newcommand{\calU}{{\cal U}}
\newcommand{\calV}{{\cal V}}
\newcommand{\calX}{{\cal X}}
\newcommand{\calY}{{\cal Y}}

\newcommand{\scrA}{\mathscr{A}}
\newcommand{\scrM}{\mathscr{M}}
\newcommand{\scrS}{\mathscr{S}}

\newcommand{\frA}{\mathfrak{A}}
\newcommand{\frM}{\mathfrak{M}}
\newcommand{\frS}{\mathfrak{S}}

\renewcommand{\proof}{\noindent{\bf Proof.\ }}

\newcommand{\lan}{\langle}
\newcommand{\ran}{\rangle}
\newcommand{\uu}{\underline}
\newcommand{\oo}{\overline}
\newcommand{\skp}{\vspace{\baselineskip}}
\newcommand{\supp}{{\rm supp}}
\newcommand{\trace}{{\rm trace}}
\newcommand{\w}{\wedge}
\newcommand{\lt}{\left}
\newcommand{\rt}{\right}
\newcommand{\pl}{\partial}
\newcommand{\abs}[1]{\lvert#1\rvert}
\newcommand{\norm}[1]{\lVert#1\rVert}
\newcommand{\mean}[1]{\langle#1\rangle}
\newcommand{\To}{\Rightarrow}
\newcommand{\til}{\widetilde}
\newcommand{\wh}{\widehat}
\newcommand{\dist}{{\rm dist}}
\newcommand{\grad}{\nabla}
\newcommand{\iy}{\infty}
\newcommand{\AddedTh}[1]{\textbf{\textcolor{Black}{#1}}}

\newcommand{\tab}{\hspace*{0.3in}}
\newcommand{\Tab}{\hspace*{1.0in}}
\newcommand{\no}{\nonumber}
\newcommand{\noi}{\noindent}
\newcommand{\txt}{\textrm}
\newcommand{\ds}{\displaystyle}
\newcommand{\RR}{\mathbb{R}}
\newcommand{\vf}{\varphi}
\newcommand{\del}{\frac{\partial}{\partial t}}
\definecolor{co}{rgb}{0.8,0,0.8}
\definecolor{gr}{gray}{0.5}
\newcommand{\gr}{\color{gr}}
\newcommand{\vp}{\varepsilon}

\newcommand{\DoPolicy}[1]{}

\title{Insulin Regimen ML-based control for T2DM patients}

\author{Mark Shifrin, Hava Siegelmann}

\maketitle
%--------------- table of contents ----------------------------------------
%

\begin{abstract}
We model individual T2DM patient blood glucose level (BGL) by stochastic process with discrete number of states mainly but not solely governed by medication regimen (e.g. insulin injections). BGL states change otherwise according to various physiological triggers which render a stochastic, statistically unknown, yet assumed to be quasi-stationary, nature of the process. 
In order to express incentive for being in desired healthy BGL we heuristically define a reward function which returns  positive values for desirable BG levels and negative values for undesirable BG levels. 
The state space consists of sufficient number of states in order to allow for memoryless assumption. This, in turn, allows to formulate Markov Decision Process (MDP), with an objective to maximize the total reward, summarized over a long run.
The probability law is found by model-based reinforcement learning (RL) and the optimal insulin treatment policy is retrieved from MDP solution. 

\end{abstract}

%put abbriviations here
%\newpage

\section{Introduction}
Diabetes mellitus type 2 (T2DM) makes up for about $90\%$ of all Diabetes cases~\cite{t2dmWHO}, with increasing rates since as early as 1960~\cite{moscou2010getting}.
The goal of therapy for T2DM is to bring the average blood glucose (BG) as close
to the normal range as possible, which can be done by various oral medications or/and insulin injections~\cite{ripsin2009management}. 
The exact dosage and frequency of the medication intakes are coordinated by blood glucose measurements normally being performed by designated glucose meters, see , e.g.,~\cite{meter}. The recent recommendations are such that T2DM patients are advised to measure their blood glucose level (BGL) at least 3 times a day~\cite{mayo}.

In this work, we address individually adopted insulin regimen control. For this purpose, we harness tools from the area of Machine Learning (ML) and stochastic control theory. In particular, we build Markov Decision Process of blood glucose level. The solution to the defined MDP is expressed by a  \textit{policy} which assigns an action according to the measured BGL. 

Note that we pose no contradiction to other known methods employed by medical care. The ML theory is agnostic to physiological processes, e.g.,  biochemical insulin impacts e.g. renal, hepatic effects of insulin resistance, effect of other possibly malfunctioning processes inside beta-cells, impairments of insulin pathways and so on.  We merely taking a different approach, modeling the insulin treatment as a controllable stochastic process.
In contrary to other known, including ML-based works, we exploit no mathematical or biological models which express chemical dependencies. 

%On the other hand,  due to the inevitable variation of blood glucose (BG) around the mean, a lower mean will result in a higher frequency of unpleasant and sometimes dangerous low BG levels.

One of the well-known difficulties related to insulin regimen control is associated with 
short and ultradian insulin secretion peaks, which, at the recent time, are not well understood,~\cite{holt2016textbook}. 
%sections 6,7,9-11,27 mostly),
The BGL is known to increase after meal intake and to decrease during fasting periods and after physical activity. 
However, the peaks render the BGL to vary according to the pattern which is hard to mathematically characterize.

In particular, patients with insulin resistance and especially those who had progressed to T2DM, have abnormal insulin oscillations, no first phase and diminished/scattered 2nd phase of insulin secretion out of beta-cells. Even more important, these oscillations may not exactly follow  glucose oscillations, like it is known to be in healthy people~\cite{holt2016textbook}.

%Each patient may have different resistance levels to insulin (e.g. lean people vs. obese people) and different peaks patterns.

In order to characterize the BGL state of a patient, we assume continuous and bounded BG axes and perform discretization by dividing the entire region into finite BG states. 
For example, between $70$ and $300$ $mg/dL$ (see, e.g.,~\cite{joslin}). Each state refers to continuous region, such that any BG in this region is assigned to the corresponding state. 
The state can be augmented by additional individually measured parameters.
In particular, we may also account for glucose absorption rate (GAT) and $HbA_{1c}$ (glycosylated haemoglobin) values.

The patient's BGL, hence, is viewed as a stochastic process which varies according to physiological dynamics and according to the patient activities. This process is the controllable subject such that the control is applied by means of medications, e.g. insulin injections (IJ). 
%Note that the physiological dynamics  is uncontrolled by the patient and is individual.
The finesse of the discretization is set in a way to assure that the next state will solely depend on the previous state and the action, i.e., IJ, if any, taken  in that state. While the BGL process is clearly continuously fluctuating through the time, the measurements are performed in discrete time slots.
Hence, we view the state of the patient at certain time marks. We are interested in transition law, i.e. transition probabilities from one state, at some time mark $t$, to the next state which the patient sees (i.e, feels) right after the next measurement which occurs at time mark $t+1$.
This law is expressed by the transition probability from a state, associated with the corresponding BGL, given the set of actions taken by the patient in that state, to the next state. 

%Since there is a difficulty to see the next state due to the patients variability, and also let us call it intra-variability. (each patient changes with long time run). 

The aforementioned state transition probabilities are \textit{learned} by a model-based reinforcement learning (RL). The input data for the RL is taken from  existing samples of measurements individually performed under effect of medication applied by physician administration. We assume that such samples exist for each individual patient. Alternatively, the individual can be assigned to a category of patients for which the \textit{initial} insulin policy is already defined. This policy will be improved henceforth and individually adjusted by applying reinforcement learning on the individual's measurements. In this document, we show results from database open for purposes of academic usage~\cite{Lichman:2013}.

The methods of ML were already  addressed in~\cite{daskalaki2016model} for the treatment of Type I diabetes Mellitis (T1DM). The paper~\cite{daskalaki2016model} proposes linearizion of MDP solution. The model is based on high dependence of insulin and glucose levels.
However, for T2DM this assumption is not necessarily true. T1DM treatment is facilitated by very frequent (e.g., every several minutes) BG measurements and actions, while T2DM measurements are normally performed several times a day. This makes T2DM control an appropriate candidate for the stochastic modeling by discrete MDP with discrete actions control.
The prediction of BGL was addressed in~\cite{georga2011glucose} and references therein, giving a stronger emphasis to T1DM. Neural Network (NN) based model is presented in~\cite{zitar2005towards}, aimed to predict BGL in T2DM patients. The training was performed on other patients for the purpose of the NN convergence. In this work, the crucial learning part is performed using individual data, with objective to individually adjust to each patient. Authors of this model suggest that convergence of NN by sample data from other patients may cause a bias and dependency on that specific data. (Note that we do use data samples from various patients for research phases of the algorithm presented in this paper, while adopting to the specific patient is fulfilled in the phase of on-line control.) Block-oriented Wiener modeling for BG prediction is done in~\cite{rollins2010free}.

Previous works which apply ML in the context of Diabetes employ a great deal of system parameters, which is especially helpful for the prediction.
Our approach is control-oriented and is designed to completely hide the complex impact of multiple parameters behind the stochastic process of the BGL, which we consider as the main indication for the insulin control purpose. That is, we allow the assumption that BGLs' stochastic process embodies in itself the impact of all relevant physical parameters (including, in most extremely simplified cases, carbohydrates intakes and physical activities), hence we rely on learning the statistical properties of the process.

%The optimization is done over a \textit{value function}, which is formulated as a summation of total reward over time.

Using the states, actions and transition laws, we formulate Markov Decision Process (MDP), which aims to maximize the average reward over time.
 %using real-time data and measurements.
For this purpose we heuristically define a mapping which transfers the  BGL state into \textit{reward}, such that a positive reward is obtained for the healthiest states and negative rewards are obtained at undesired BG levels. We also introduce non-linear components which aim to fine for being in dangerous states, e.g., overly high BGL or hypoglycemia.  

%The obtained policy will provide a better control and faster control adjustment for patients and the medical stuff.

%The model is intended to fit all types, T1DM, T2DM, and may be other rare types.

We formally define the MDP and describe the RL algorithm in the next section.

\section{The probabilistic model}
We start by defining the state space, next we define the reward function, the reward functional and the action space.
\subsection{Definition of spaces}
\subsubsection{The measurement space}
We consider blood glucose interval of $gl_{min}$ to $gl_{max}$, where the lower bound means an extremely low BGL (possibly corresponding to hypoglycemia) while the upper bound means extremely high level of glucose in blood.
The BGL finesse coefficient (BFC) defines the size of the interval which is mapped to the single BG state.
For example for $BFC=1$ and $gl_{min}=70$$mg/dL$, the lowest interval of $[70,71)$ corresponds to the lowest \textit{state}. The set of all possible BGL forms the space:
\[
\LL=\{1,\cdots,L\},\;\;L=\frac{gl_{max}-gl_{min}}{BFC}.
\]
Denote the \textit{BGL-state} by $gl$, where $gl\in\LL$,

We allow to augment state space by additional measured parameters, e.g., glucose absorption rate (GAT),  Similarly to BGL, we perform discretization for GAT by defining the bounds of the GAT axes $ga_{min}$ to $ga_{max}$.
The GAT finesse coefficient (GFC) is used in order to define the GAT-state: 
\[
\GG=\{1,\cdots,G\},\;\;G=\frac{ga_{max}-ga_{min}}{GFC}.
\]

The \textit{measurement space} $\MM$ defines the set of all possible measurements of the patient.
In the case both BGL and GAT can be measured at all times we have
$\MM=\LL\times\GG$.
% $\mathcal{M}=\mathcal{L}\times\mathcal{G}\times\mathcal{T}$.
In the simplest scenario where the only component of the measurement space is BGL, we merely have
$\MM=\LL$.

\subsubsection{The daytime space}
We differentiate the BGL process instantiations during the day at different \textit{measurement points}. 
These points are expressed by \textit{daytime space}, denoted by
\[
\TT=\{\tau_1,\cdots,\tau_T\},
\]
where $T$ is the number of measurements per day. For simplicity we assume this number is constant.
\begin{example}
	Consider approximately constant measurement timings, corresponding to  morning, noon, afternoon, evening, night.
	
	Then, $T=5$ and $\{\tau_1,\tau_2,\tau_3,\tau_4,\tau_5\}=\{\text{morning, noon, afternoon, evening, night}\}$.
	\label{ex1}
\end{example} 

The detailed time is denoted by a couple $\{t,\tau\}$. 
In this sense $t\in\{0,1,\cdots\}$ represents the date, while $\tau\in\TT$ represents the time during the day, corresponding to one of the measurements. 
The absolute measurement points can be counted by denoting $\iota=t\cdot T+\tau$.
At any $\iota$ we consider $ms(\iota)\in\MM$.
%We will omit $t$ when we will refer to an arbitrary date.

\subsubsection{Carbohydrates intake history space}
The carbohydrates finesse coefficient (CFC) is used in order to define the quantity of carbohydrates consumed at some time mark by a patient: 
\[
\CC=\{1,\cdots,C\},\;\;C=\frac{ch_{max}-ch_{min}}{CFC}.
\] 
In order to wisely incorporate the Carbohydrates intake history (CIH) we will assume that at each $\iota$ last $H$ intakes of carbohydrates are relevant for the insulin treatment. That is, at any measurement point $\iota$ we care of all measurement in the range $[\iota,\iota-H]$, while measurements of older points are presumed of no longer having impact on the individual. 
The CIH at time $\iota$ is denoted by $CH(\iota)=\{ch_\iota\}$, where $ch_i\in\CC,\;\forall i\in[\iota,\iota-H]$.

See that in example~\ref{ex1}, for $H=3$, at time $\{t,\text{noon}\}$  we have  $CH(\{t,\text{noon}\})=\{ch\{t,noon\},ch\{t,morning\},ch\{t-1,night\},ch\{t-1,evening\}\}$, while $CH(\{t,\text{afternoon}\})=\{ch\{t,afternoon\},ch\{t,noon\},ch\{t,morning\},ch\{t-1,night\}\}$.

Such a structure allows for impact of the past carbohydrate intakes on one hand and preserves memoryless property on th other hand. That is, $CH(\iota)$ depends only on $CH(\iota-1)$ and $ch(\iota)$.

\subsubsection{Physical activity history space}
We assume physical activities have a prolonged effect, similarly to CIH.  The measurements of energy spent for a given activity can be done in metabolic equivalents, or METs, see, e.g., the table of translation of activity to METs in~\cite{phyHVD}. The physical activity finesse coefficient (PFC) is used in order to define the number of METs corresponding to the energy spent by an individual at some measurement (or in a period between two consecutive measurements): 
\[
\PP\HH=\{1,\cdots,PH\},\;\;C=\frac{ph_{max}-ph_{min}}{PFC}.
\] 
To define the Physical activity history (PAH), denote at time $\iota$ the set $PH(\iota)=\{ph_\iota\}$, where $ph_i\in\PP\HH,\;\forall i\in[\iota,\iota-Y]$, and $Y$ is the size of the relevant history.

\subsubsection{Patient activity space}
We define the patient activity (PA) space by 
\[
\PP\AAA=\CC\HH\times\PP\HH
\] 
At any $\iota$ we consider $pa(\iota)\in\PP\AAA$. We reason we differ between PA space and the measurement space is summarized as follows
\begin{itemize}
	\item $ph$ is directly known by a patient and can be foreseen by conducted activities. In contrary, $ms$ is assumed to be purely stochastic and can be only measured.
	\item The insulin medication is applied in order to control the $ms$, i.e., the BGL, while $pa$ may only effect the dosage.
	\item We aim to apply optimal control for given $ms$ and ,optionally, for a given $pa$. The $pa$ can be seen as the secondary means of control, however the maintaining certain level of activities is not a part of the insulin policy (see definition below).
\end{itemize}

\subsubsection{Overall state space definition}
We now ready to define the state space and the state.
Denote
\[
\SSS=\PP\AAA\times\MM\times\TT
\]
Denote by $s_{\iota}\in\SSS$,  a state at absolute measurement point $\iota$. %, such that $s_{(t)}\in\SSS$.
That is, the state consists of BGL and other measurements according to definition of $\MM$, patient activities history according to definition of $\PP\AAA$ and daytime of $\iota$, according to definition of $\TT$.

Note that in the simplest scenario where the only component of the measurement space is BGL and no patient activities are accounted for, we merely have
$\SSS=\LL\times\TT$, and $s=\{gl,\tau\}$, where $gl\in\LL$ and $\tau\in\TT$. 

Also note, that even though $\iota$ already carries the information about $\tau$, we will still sometimes explicitly specify $\tau$, for the clarity.

\subsection{Reward function}
Denote the optimal (the most healthy) BGL by $gl_h$.
Define reward function $r(s_{(t,\tau)})$ which maps the BGL at $s$ at some given time $\{t,\tau\}$ to the value which \textit{quantifies the quality of the healthiness} of a patient with BGL at $s_{(t,\tau)}$.
The linear reward component, normalized to $1$, is given by 
\[
r^l(s)=r^l(g)=1-\frac{|gl-gl_h|}{gl_{1}-gl_{L}}
\] 
Define a subset of dangerous states $\LL_{hyp}\in\LL$ of being close to hypoglycemia by $\LL_{hyp}=\{1,\cdots,L^c\},\;\;L^c<L$. 
Normally, $L^c$ is small.  It defines the number of states with critically low BGL.
Define the non-linear reward component which fines for being close to hypoglycemia
\[
r^n(s)=r^n(g)=\bI_{gl<gl_{L^c}}\cdot[1-\frac{gl-gl_{L^c}}{gl_{L^c}}],
\] 
where $\bI$ is the standard indicator function.
The joint instantaneous reward is given by \[r(g)=r^l(g)-r^n(g)\].

%We would like to discover/adjust an appropriate insulin regimen for T1DM or T2DM. (only basal, short-acting, mixed, timings of taking tablets/injections), I understand that there are various kinds of treatments (soluble insulin, RAI, combinations, etc.), with varying activation times and acting times.

\subsection{Action Space}

The action space consists of set of possible insulin related medical actions. We allow for finite number of treatments, in particular, the most known of them - short-acting (Actrapid-like) insulin preparations, intermediate-acting (NPH-like) and long-acting (Ultratard-like) insulin preparations.
Denote the sets of possible dosages 
\begin{align*}
\AAA^s=\{a^s_1,\cdots,a^s_{|\AAA^s|}\},\;\;\AAA^i=\{a^i_1,\cdots,a^i_{|\AAA^i|}\},\;\;
\AAA^l=\{a^l_1,\cdots,a^l_{|\AAA^l|}\},
\end{align*}
corresponding the short-acting, intermediate-acting and long-acting insulin medications. Note that the first values of the spaces $a^s_1,a^i_1,a^l_1$ are  equal to $0$,  meaning no mediation of that type is given. 
As long as the medications can be administered concurrently we assume that the action space is given by
\[
\AAA=\AAA^s\times\AAA^i\times\AAA^l
\]
At each state $s\in\SSS$, an action $a\in\AAA$ is chosen. At time mark $\iota$, $a_{(\iota)}$ is a triplet $\{a^s_{(\iota)},a^i_{(\iota)},a^l_{(\iota)}\}$.

 %(normally, set of all possible medications,  patient-initiated activities which impact, e.g. sport exercise) which will be mapped to results (that is, action outcomes- blood glucose vs. insulin measurements). The good results will be associated with so-called positive reward, while the bad results (e.g. some  hypoglycemia was caused, no target level achieved) will be associated with negative rewards. %These rewards we should assume and quantify in advanced.

\subsection{Transition Probabilities}
The transitions between states occurs at every absolute measurement point according to transition probabilities given by
$p(s'_{(\iota+1)}|s_{(\iota)},a_{(\iota)})$. That is, the probability of getting to the next state $s'$, given that the previous state was $s$, and action $a$ was taken.
Clearly,
\[
\sum_{s'}p(s'_{(\iota+1)}|s_{(\iota)},a_{(\iota)})=1,
\]
where the summation is over all possible states at $\iota+1$.

Clearly, these probabilities are unknown  a priori. Hence, the  objective is to statistically learn them using a patient's measurements, \textit{under previously applied insulin regime}. This regime could be applied according to doctor's purely medical (i.e., non-mathematical) considerations. Once these probabilities are learned, the new, optimal insulin policy is derived. 
%The learning of the transition probabilities and procedure of converging to the optimal function is be done iteratively.

%Next, the action and the previous state will be also mapped to the new state  which will be measured after the common effect of
%(passed time period)vs.(meal taken/ingested)vs.(level of glucose vs insulin in blood).

\subsection{Total reward functional}
Define the total discounted reward, with positive discount factor $\gamma<1$.
\begin{equation}
J=\EE\sum_{\iota=0}^\infty \gamma^\iota r(s_{(\iota)}),
\label{piux1}
\end{equation}

The discount factor $\gamma$ has two important interpretations.
The first one follows from the known analytical property which allows the sum in~\eqref{piux1} to converge. 
The second one advocates the reasoning of "feeling better now" is more important than "feeling better in a distant future".
Therefore, BGL rewards for the distant time marks are discounted with geometrically decreasing discount multiplicative. 
The alternative approach, which values equally the feeling throughout the day employs the following  definition
\begin{equation}
J=\EE\sum_{t=0}^\infty\sum_{\tau=\tau_1}^{\tau_T} \gamma^t r(s_{(t,\tau)})=\EE\sum_{t=0}^\infty\gamma^t\sum_{\tau=\tau_1}^{\tau_T}  r(s_{(t,\tau)}).
\label{piux5}
\end{equation}

%\section{Related work }

\section{MDP and RL solution}
\subsection{The Bellman equation}
Write the equation~\eqref{piux1}, for the given initial state $s_{(0)}$ as follows
\begin{equation}
J^\pi(s_{(0)})=\EE\sum_{t=0}^\infty \gamma^tr(s_{(t)})=r(s_{(0)})+\EE\sum_{t=1}^\infty \gamma^tr(s_{(t)})
\label{piux2}
\end{equation}
Denote by $\pi$ a policy which selects action at each state. 
%The objective is to find an optimal policy,.
State $s(1)$ is set at time mark $t=1$, according to the transition probability from state $s_{(0)}$, given action 
$a^\pi(s_{(0)})=\{a^{\pi,s}(s_{(0)}),a^{\pi,i}(s_{(0)}),a^{\pi,l}(s_{(0)})\}$, which is chosen according to $\pi$.  

Define the \textit{value function} given an initial state:
\begin{equation}
V(s_{(0)})=\max_\pi J^\pi(s_{(0)})
\end{equation}
$V(s_{(0)})$ indicates the optimal total reward obtained for the BGL throughout time, when starting BGL was as in  $s_{(0)}$.
We use next the dynamic programming (DP) principle. Expand~\eqref{piux2} as follows
\begin{equation}
J^\pi(s_{(0)})=r(s_{(0)})+\EE\sum_{t=1}^\infty \gamma^tr(s_{(t)})=r(s_{(0)})+\sum_{s_{1}}p(s_{1}|a(s_0),s_0)J^\pi(s_{(1)})
\label{piux3}
\end{equation}
Applying DP on the value function gives the Bellman equation (see e.g.,~\cite{bertsekas1995dynamic})
\begin{equation}
V(s_{(0)})=r(s_{(0)})+\max_{a(s_0)}\sum_{s_{1}}p(s_{1}|a(s_0),s_0)V(s_{(1)})
\label{piux4}
\end{equation}
%The policy  can be retrieved right from the optimized avg. reward. The learning will always continue in order to compensate for the personal changes (transition probabilities may change, insulin action may change within months or so).
Denote $V$ the vector of all $V(s)$. The Bellman equation~\eqref{piux4} is solved by the value function, a result of a mapping from the state space to the space of the total discounted reward, which is complete metric space, denote it by $\mathcal{V}$:
\[
V:\SSS\mapsto\mathcal{V},
\]
equipped with a suitable metric $\Arrowvert V_a-V_b\Arrowvert$. Observe that~\eqref{piux4} can be written with operator $\T$, that is, 
$V=\T V$.
By a well-known result (see, e.g.,~\cite{bertsekas1995dynamic}), operator $\T$ is strict contraction (i.e., $\alpha\Arrowvert V_a-V_b\Arrowvert<\Arrowvert V_a-V_b\Arrowvert$, for some $0<\alpha<1$). Therefore, $\T$ has a fixed point (see, e.g.~\cite[Theorem V.18]{reed1980methods}).
This fixed point is $V$ and it constitutes the unique solution to the Bellman equation~\eqref{piux4}. The reasoning above means $V$ is found by a well-known method of \textit{value iteration} (see, again,~\cite{bertsekas1995dynamic}). That is,  $\T$ is iteratively applied, starting with initial values of some $U\in\mathcal{V}$, till the arbitrarily close convergence to the fixed point $V$.

\subsection{Algorithm formulation}%\subsection{Learning the transition probabilities}
Observe that~\eqref{piux4} can only be solved by using transition probabilities, as were defined in previous section.
These probabilities constitute the dynamics of the stochastic process of BGL and the response to the actions, i.e. the insulin preparations.
The straightforward approach to calculate them is by mere statistical learning over the existing vector of samples of measurements of BGL paired with the corresponding actions. %This initial vector is used to find the initial policy. 

We formulate the complete \textit{Insulin Optimization Policy Algorithm} next. See Figure~\ref{fig1} below for the complete formulation. 
The \textit{phase A} is the initialization of the vector of statistics of $\nu(s',a, s)$ which counts all occurrences of being in state $s$, acting by applying $a$ and passing to the next state $s'$.
The \textit{phase B} comprises the learning phase where acting is done by initial policy $\pi^0$, which normally taken from the same samples, and stems from doctor's heuristic consideration treating  the individual patient. The output of this phase is the transition probabilities.
These probabilities are exploited in the \textit{phase C} to perform the value iteration and to retrieve the optimal policy $\pi^*$. The number of iterations depends on the size of the vector $V$ and is heuristically set, in order to achieve the desired level of convergence.
Finally, \textit{phase D} is the exploitation part, which exploits the latest found optimal policy $\pi$. The phase keeps tracking the transition occurrences and updates the policy every $M$ transitions. 

The final phase reflects possible changes in the BGL process properties through the time.
 
\begin{figure}[h!] 
\fbox{\begin{minipage}{0.85\textwidth}
{%\small
	{\it \textbf{Insulin Optimization Policy Algorithm}} \\{\it\hspace*{1ex} A. Initialization}
 				\begin{enumerate}
 					\item Initialize statistical data $\nu(s',a, s)=0$, for all triplets $\{s',a,s\}$ %$r(\hat s',\hat a,\hat s)=0$.
 					\item Initialize the value function $V(s)$ for all initial states $s$.
 				%	\item Run $\scrM$ with $\pi_{1}^0$ for $N_0$ time marks. %Each visit to $s'$, find $f_1(s')=\bar s'$. Use $\hat s'\sim\bar s'$ to update by sampling $\hat p(\hat s'|\hat a,\hat s)$ and $r(\hat s',\hat a,\hat s)$.
 					%\item Calculate $\pi_B^1$ by finding $\hat V_0$ from solution to the Bellman equation (performing value iteration), using sampled probabilities and sampled reward found in step $2$.
 				\end{enumerate}}
 				{%\small
 					{\it B. Learning from the training sequence of length $N$}
 					\begin{enumerate}
 						\item set $n=0$
 						\item Visit $s=s_n$, see the action $a=a_{n}$, $a\in\pi^0$, taken in that state and the next state $s'=s_{n+1}$. 
 						Update
 						\(
 						\nu(s',a, s)=\nu(s',a, s)+1
 						\)
 						\item Increment $n$, if $n<N$ go to 1.
 						\item Calculate \(p(s'|a,s)=\frac{\nu(s',a, s)}{\sum_{s'}\nu(s',a, s)}\)
 						\end{enumerate}
 					}
{%\small
	{\it C. Value function and policy calculation}
	\begin{enumerate}
		\item Use $p(s'|a,s)$ to perform value iteration till desired level of convergence of $V$.
		\item Apply the $max$ operator at each state $s$ to find the best action. Set $argmax_a\in\pi^*$, the optimal policy. 
	\end{enumerate}
}
{{\it D. Policy on-line tracking and calibration}
\begin{enumerate}
	\item Initialize $M$, set $n=0$.
	\item Visit $s=s_n$, see the action $a=a_{n}$, $a\in\pi^*$, taken in that state and the next state $s'=s_{n+1}$. 
	Update
	\(
	\nu(s',a, s)=\nu(s',a, s)+1
	\)
	\item Increment $n$, if $n<M$ go to 1.
	\item Calculate new $p(s'|a,s)$
	\item Use new $p(s'|a,s)$ to perform again value iteration
	\item Update the new $\pi^*$.
\end{enumerate}
}

\end{minipage}}
\label{fig1}
  \end{figure}

\section{Optimal policy result}
In this section we use data from~\cite{Lichman:2013}, in order to demonstrate the activation of the algorithm described in the previous section.
Both examples presented in this section are solely based on short vectors of several hundreds of measurements of BGL. Anything else, e.g., carbohydrates intakes and physical activities are presumed to be embodied in the BGL process.
Clearly, the obtained policy is only a presumable outcome, as the learning was done over short-sized vectors. While the BGL (simply denoted as Glucose in the graphs below) scale is presented starting from minimal measurement and ending by maximal BGL measurement, not at all points in the BGL scale the policy any policy is indicated.
The reason for this stems from the limited data size. In particular, the action outcome cannot be learned in the case no occurrence of any action for that specific BGL was ever recorded.
For this reason the preliminary policy $\pi^0$ should be carefully selected. The best anticipated practice would be having an assessment of a particular individual and assigning them to the \textit{specific category of policies}. Once the successful assignment is done, the correction is expected to be effective even if only limited period of individualized measurements is offered. 
We further elaborate in the Future Work section.

\begin{figure}
	\center
	\includegraphics[scale=0.35]{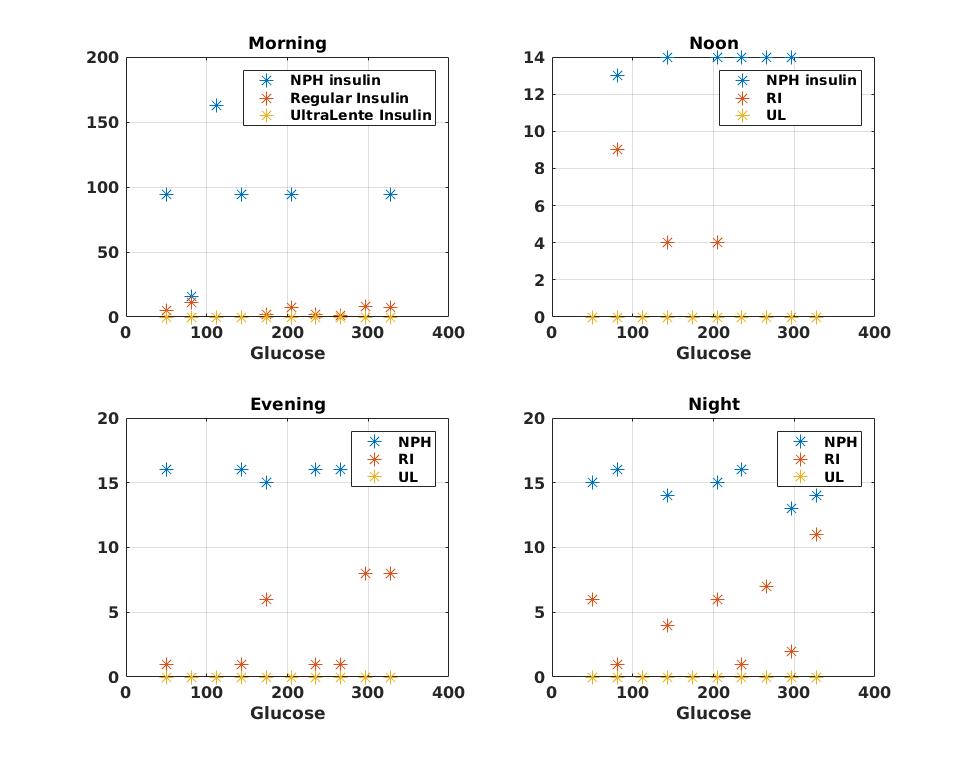}
	\vspace*{-5pt}\caption{Policy example 1, 943 measurement points. }\label{fig:32} \vspace*{-15pt}
\end{figure}

\begin{figure}
	\center
	\includegraphics[scale=0.35]{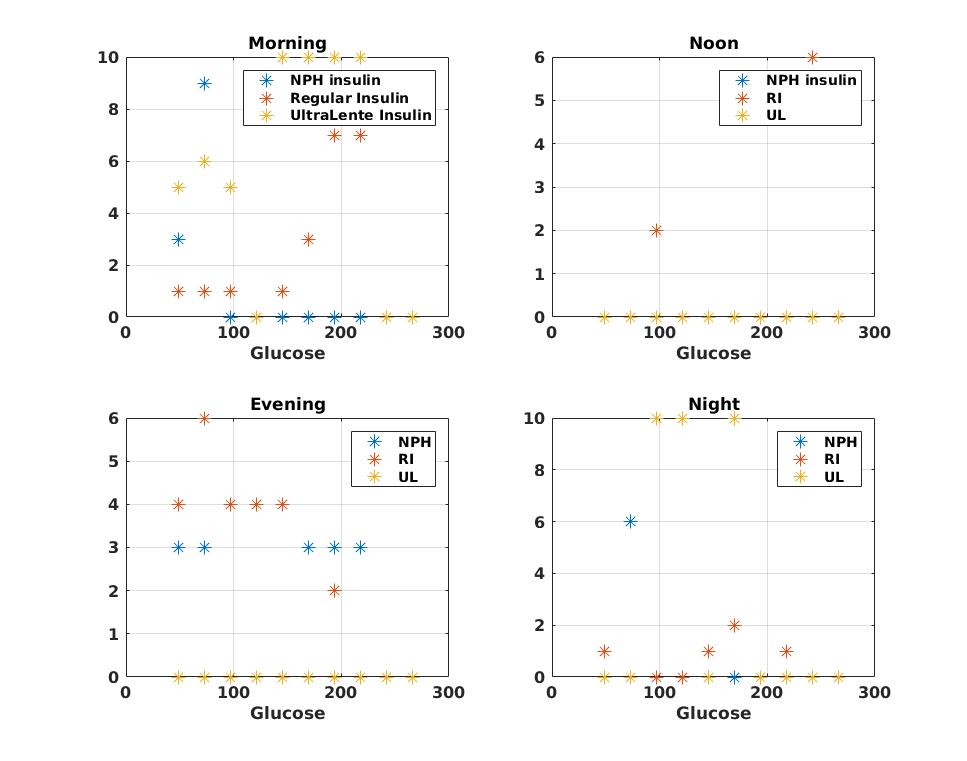}
	\vspace*{-5pt}\caption{Policy example 2, 340 measurement points. }\label{fig:33} \vspace*{-15pt}
\end{figure}

%\appendix
\section{Future Work}\label{sec_f}
The future work is aimed at two directions.
\begin{itemize}
	\item Producing a set of categories of individuals. The individuals would be assigned into the finite number of \textit{policy sets}. Each set will contain patients with resembling (but clearly not similar) insulin regime. The patients may also have other resembling parameters, e.g., body weight, age, etc. The categorizing process will be done by both heuristic assignment and correlation study. 
	\item The second direction relates tot the pure ML approach and is directed towards coping with the size of of the state-space. For this purpose, approach from approximate MDP (AMDP) will be introduced. See, e.g., methods described in~\cite[Chapter 7]{powell2007approximate}. 
\end{itemize}

\bibliographystyle{abbrv} \bibliography{insreg}

\begin{thebibliography}{10}

\bibitem{bertsekas1995dynamic}
D.~Bertsekas.
\newblock {\em Dynamic programming and optimal control}, volume~2.
\newblock Athena Scientific Belmont, MA, 1995.

\bibitem{daskalaki2016model}
E.~Daskalaki, P.~Diem, and S.~G. Mougiakakou.
\newblock Model-free machine learning in biomedicine: Feasibility study in type
  1 diabetes.
\newblock {\em PloS one}, 11(7):e0158722, 2016.

\bibitem{georga2011glucose}
E.~I. Georga, V.~C. Protopappas, and D.~I. Fotiadis.
\newblock Glucose prediction in type 1 and type 2 diabetic patients using data
  driven techniques.
\newblock In {\em Knowledge-oriented applications in data mining}. InTech,
  2011.

\bibitem{phyHVD}
{Harvard School of Public Health}.
\newblock {Measuring Physical Activity}.
\newblock
  \url{https://www.hsph.harvard.edu/nutritionsource/mets-activity-table/},
  2017.

\bibitem{holt2016textbook}
R.~I. Holt, C.~Cockram, A.~Flyvbjerg, and B.~J. Goldstein.
\newblock {\em Textbook of diabetes}.
\newblock John Wiley \& Sons, 2016.

\bibitem{joslin}
{Joslin Diabetes Center}.
\newblock {Blood glucose chart}.
\newblock
  \url{http://www.joslin.org/info/Goals-for-Blood-Glucose-Control.html}, 2017.

\bibitem{Lichman:2013}
{M. Lichman, University of California, Irvine, School of Information and
  Computer Sciences}.
\newblock {UCI} machine learning repository.
\newblock
  \url{https://archive.ics.uci.edu/ml/datasets/Diabetes+130-US+hospitals+for+years+1999-2008},
  2013.

\bibitem{mayo}
{Mayo Clinic}.
\newblock {Blood sugar testing: Why, when and how}.
\newblock
  \url{http://www.mayoclinic.org/diseases-conditions/diabetes/in-depth/blood-sugar/art-20046628},
  2017.

\bibitem{moscou2010getting}
S.~Moscou.
\newblock Getting the word out: advocacy, social marketing, and policy
  development and enforcement.
\newblock {\em Public Health Nursing}, page 285, 2010.

\bibitem{powell2007approximate}
W.~B. Powell.
\newblock {\em Approximate Dynamic Programming: Solving the curses of
  dimensionality}, volume 703.
\newblock John Wiley \& Sons, 2007.

\bibitem{reed1980methods}
M.~Reed and B.~Simon.
\newblock {\em Methods of modern mathematical physics: Functional analysis},
  volume~1.
\newblock Gulf Professional Publishing, 1980.

\bibitem{ripsin2009management}
C.~M. Ripsin, H.~Kang, R.~J. Urban, et~al.
\newblock Management of blood glucose in type 2 diabetes mellitus.
\newblock {\em Am Fam Physician}, 79(1):29--36, 2009.

\bibitem{rollins2010free}
D.~K. Rollins, N.~Bhandari, J.~Kleinedler, K.~Kotz, A.~Strohbehn, L.~Boland,
  M.~Murphy, D.~Andre, N.~Vyas, G.~Welk, et~al.
\newblock Free-living inferential modeling of blood glucose level using only
  noninvasive inputs.
\newblock {\em Journal of process control}, 20(1):95--107, 2010.

\bibitem{meter}
{U.S. National Library of Medicine}.
\newblock {Blood Glucose Self-Monitoring}.
\newblock \url{https://meshb.nlm.nih.gov/record/ui?ui=D015190}, 2017.

\bibitem{t2dmWHO}
{World Health Organization}.
\newblock {Diabetes facts list}.
\newblock
  \url{https://web.archive.org/web/20130826174444/http://www.who.int/mediacentre/factsheets/fs312/en/},
  2013.

\bibitem{zitar2005towards}
R.~A. Zitar and A.~Al-Jabali.
\newblock Towards neural network model for insulin/glucose in diabetics-ii.
\newblock {\em Informatica}, 29(2), 2005.

\end{thebibliography}
% that's all folks
\end{document}